\documentclass[11pt,a4paper]{article}
\usepackage{amsfonts}
\usepackage{amssymb}
\usepackage{graphicx}
\usepackage{epstopdf}
\usepackage{cite}
\usepackage{color}
\usepackage{verbatim}
\usepackage{amsmath}
\usepackage{wasysym}
\usepackage[thickspace,amssymb]{SIunits}
\usepackage[T1]{fontenc}
\usepackage[utf8]{inputenc}
\usepackage{jnm}
\usepackage{setspace}

\def\mycite#1{~(\citen{#1})}

\title{Towards a Radio-guided Surgery with $\beta^-$ Decays: Uptake of a somatostatin analogue (DOTATOC) in Meningioma and High Grade Glioma. }

\author{ Francesco~Collamati$^{1,2}$, Alessandra~Pepe$^3$, Fabio~Bellini$^{1,2}$, Valerio~Bocci$^2$, Marta~Cremonesi $^4$, Erika~De Lucia$^{5}$, Mahila~Ferrari$^4$, Paola~M. Frallicciardi$^{2,6}$, Chiara M. Grana$^4$, Michela~Marafini$^{2,6}$, Ilaria~Mattei$^{5,7}$, Silvio~Morganti$^2$, Vincenzo~Patera$^{2,3}$, Luca~Piersanti$^{2,3}$, Luigi~Recchia$^2$, Andrea~Russomando$^{1,2,8}$, Alessio~Sarti$^{5,3}$, Adalberto~Sciubba$^{2,3}$, Martina~Senzacqua$^{1,2}$, Elena~Solfaroli~Camillocci$^8$, Cecilia~Voena$^{2}$, Riccardo~Faccini$^{1,2}$
\\\\
{$^1$} Dipartimento di Fisica, Sapienza Universit\`a di Roma, Roma, Italy; 
{$^2$} INFN Sezione di Roma, Roma, Italy; 
{$^3$} Dipartimento di Scienze di Base e Applicate per l'Ingegneria, Sapienza Universit\`a di Roma,  Roma, Italy;
{$^4$} Istituto Europeo di Oncologia, Milano, Italy; 
{$^5$} Laboratori Nazionali di Frascati dell'INFN, Frascati, Italy; 
{$^6$} Museo Storico della Fisica e Centro Studi e Ricerche ``E.~Fermi'', Roma, Italy;
{$^7$} Dipartimento di Fisica, Universit\`a RomaTre, Roma, Italy;
{$^8$} Center for Life Nano Science@Sapienza, Istituto Italiano di Tecnologia, Roma, Italy.}

\begin{document}
   \maketitle
   
{\it  Corresponding author} : R. Faccini, Dip. Fisica, Universita' di Roma "La Sapienza", P.le A. Moro 2, 00185, Rome, Italy. email: riccardo.faccini@roma1.infn.it, tel: +39 0649914798, fax: +39 06 4957697.

{\it  First author} : F. Collamati, Ph.D. student, Dip. Fisica, Universita' di Roma "La Sapienza", P.le A. Moro 2, 00185, Rome, Italy. email: francesco.collamati@roma1.infn.it, tel: +39 0649914998, fex: +39 06 4957697

{\it Word Count}:  3600 words

The study was financed by the research funds of the Universit\`a di Roma "La Sapienza"

\vspace{2cm}
{\it Short Running Title}: DOTATOC Uptake In Meningioma and Glioma

\section*{Abstract}
A novel radio guided surgery (RGS) technique for cerebral tumors using $\beta^-$ radiation is being developed. Checking the availability of a radio-tracer that can deliver a $\beta^-$ emitter to the tumor is a fundamental step in the deployment of such technique. This paper reports a study of the uptake of $^{90}$Y-labeled [1,4,7,10-tetraazacyclododecane-N,N$^\prime$,N$^{\prime\prime}$,N$^{\prime\prime\prime}$-tetraacetic acid0-D-Phe1,Tyr3]octreotide (DOTATOC) in the meningioma and the high grade glioma (HGG) and a feasibility study of the  RGS technique in these cases. Such estimates are performed assuming the use of a  $\beta^-$ probe with a sensitive area of 2.55~mm radius, that is under development, to detect 0.1ml residuals.
{\bf Methods:} the uptake and the background from healthy tissues were estimated on  $^{68}$Ga-DOTATOC PET scans of 11 meningioma and 12 HGG patients. A dedicated statistical analysis of the DICOM images was developed and validated. The feasibility study was performed by means of a full simulation of the emission and detection of the radiation, accounting for the measured uptake and background rate.
{\bf Results:}   all meningioma patients but one with an atypical extra-cranial tumor showed a very high uptake of DOTATOC. In terms of feasibility of the RGS technique, we estimated that  by administering 3~MBq/kg of radio-tracer, the time needed to detect a 0.1~ml remnant at 95\% C.L. is smaller than 1~s. Actually, to achieve a detection time of 1~s the required activities to administer are as low as 0.2-0.5 MBq/kg in a large fraction of the patients.
In case of HGGs, the uptake is lower, but the tumor-to-non-tumor ratio is higher than four, which implies that it can still be effective for RGS. It was estimated that by administering 3~MBq of radio-tracer, the time needed to detect a 0.1~ml remnant at 95\% C.L. is smaller than 6~s with the exception of the only oligodendrioma in the sample.
{\bf Conclusion:}
 The uptake of $^{90}$Y-DOTATOC in meningioma is high in all studied  patients. As far as HGG is concerned, albeit the uptake is significantly worse, it is still acceptable for RGS, in particular if further R\&D is made to improve the performances of the $\beta^-$ probe.

{\bf Keywords}:Radio Guided Surgery, somatostatin analogue, Meningioma, High Grade Glioma.
\newpage
\doublespacing
\section*{Introduction}
The radio-guided surgery (RGS) is a technique that facilitates the surgeon 
when evaluating the completeness of the tumoral lesion resection, 
while minimizing the amount of healthy tissue removed\mycite{RadioGuided}.
The impact of the RGS on the surgical management of cancer patients 
includes providing the surgeon with vital and real-time information
regarding the location and the extent of the lesion, 
as well as assessing surgical resection margins.
The technique makes use of a radio-labelled tracer, 
preferentially uptaken by the tumor  
to discriminate the cancerous tissue from the healthy organs, 
and a probe (for a review see\mycite{IntrProbes}), 
sensitive to the emission released by the tracer, 
to identify in real time the targeted tumor focus. 
The radio-pharmaceutical is administered to the patient 
before surgery. 

Current clinical applications of the RGS are: 
radio-immuno-guided surgery (RIGS) for colon cancer\mycite{colon,colon2}, complete 
sentinel-node mapping for malignant melanoma\mycite{melanoma} and breast cancer\mycite{breast,breast2}, 
and detection of parathyroid adenoma\mycite{adenoma} and bone tumors 
(such as osteoid osteoma).
%RF added 
 There are also clinical studies on the application to neuro-endocrine tumors (NET)\mycite{Baum,Baum2}.

 Established methods make use of a combination of
a $\gamma$-emitting tracer with a
$\gamma$ radiation detecting probe (see for instance Refs\mycite{GammaProbes,GammaReview} and references therein).
Since $\gamma$ radiation can traverse large amounts of tissue,
any uptake of the tracer in nearby healthy tissue
represents a non-negligible background, 
often preventing the usage of this technique. 

To overcome these limits and extend the range of applicability of RGS, it was suggested \mycite{SciRep} to use   pure $\beta^{-}$-emitting radio-isotopes instead of the $\gamma$-emitting tracers. $\beta^-$ radiation indeed  penetrates only a few millimetres of tissue with essentially no $\gamma$ contamination, since the \textit{bremsstrahlung} contribution, that has a 0.1\% emission probability, can be considered negligible.  This novel approach allows to develop a handy and compact probe which, 
detecting electrons and operating with low radiation background, 
provides a clearer delineation of margins of the lesioned tissue. It also requires a smaller activity 
to detect tumor remnants compared to traditional RGS approaches so that, due to the lower absorbed dose\mycite{SciRep} and the short range of electrons, the radiation exposure of the medical personnel becomes almost negligible.

For this novel technique to be applicable, tracers capable to deliver pure $\beta^-$ emitting radio-nuclides are needed. Among the existing tracers, those used for molecular radio-therapy, in particular the Peptide Receptor Radionuclide Therapy (PRRT) with $^{90}$Y, are the first candidates to be considered. Furthermore, neurosurgery is one of the fields that would profit most from RGS, given the importance of complete lesion removal. Finally, pre-surgical imaging is scarcely effective for the search for remnants  because the operative field changes substantially during the brain tumor resection. This makes $^{90}$Y-labeled [1,4,7,10-tetraazacyclododecane-N,N$^\prime$,N$^{\prime\prime}$,N$^{\prime\prime\prime}$-tetraacetic acid0-D-Phe1,Tyr3]octreotide (DOTATOC) an optimal candidate tracer for a first application of RGS with $\beta^-$ radio-nuclides in brain tumors: meningioma has long been known to have an important uptake of DOTATOC\mycite{MeningiomaUptake,MeningiomaUptake2,MeningiomaUptake3,MeningiomaUptake4} and there are recently studies showing that also HGG presents receptors to somatostatin\mycite{glioblastomaUptake}.

Existing studies\mycite{MeningiomaUptake} focus on the characteristics of the uptake needed for molecular radio-therapy. This paper studies instead, in the case of meningioma and HGG, the aspects that are critical in the case of RGS, when the administered activities are significantly lower and the required tumor-to-non-tumor ratio (TNR)  is smaller.

\section*{MATERIALS AND METHODS}
\subsection*{Estimate of The DOTATOC Uptake}

The first goal of this study was to assess the uptake of $^{90}$Y-DOTATOC in meningioma, HGG and nearby healthy tissue. To this aim, we have examined PET scans with  $^{68}$Ga-DOTATOC of the two diseases\mycite{Ga-DOTATOC}, with the reasonable assumption that Ga and Y are chemically similar enough not to alter the kinetics of the tracer.
We have retrospectively analyzed, with the AMIDE software, %\mycite{AMIDE},
 DICOM images of 11 patients affected by meningioma and 12 by HGG and estimated the mean tumor uptake and the uptake of the nearby healthy tissues (the "non-tumor" estimate needed to compute the tumor-to-non-tumor ratio, TNR). All patients gave their written informed consensus to clinical research.  A summary of the characteristics of the meningioma and HGG patients are in Tabs.~\ref{tab:meningioma} and~\ref{tab:glioblastoma} respectively.

\begin{figure}[bth!]
\begin{center}
\includegraphics [width = 174 mm]  {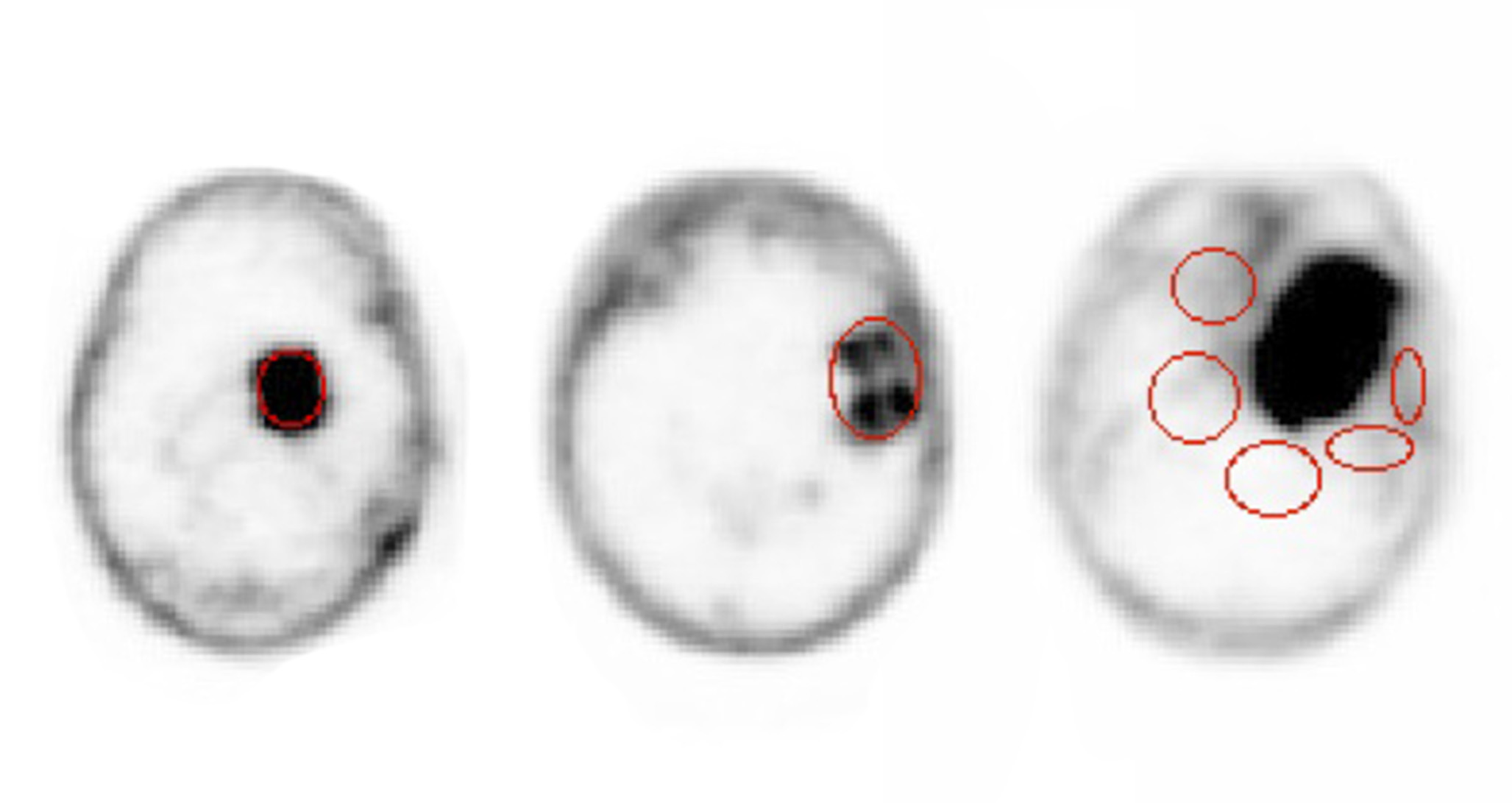}
\caption{Example of ROI definition for meningioma (left), HGG (center), and non-tumor (right). }
\label{fig:ROI}
\end{center}
\end{figure}

To estimate the mean tumor uptake, for each meningioma patient we considered several slices of the PET scan and in each of it a region of interest (ROI), containing only lesioned tissue on the basis of the CT scans, was defined. In presence of  several lesions, each one was considered separately (see Fig.~\ref{fig:ROI}  on the left for an example). In the case of the HGG, instead, there was often a  necrotic area in between  uptaking tissues (center of Fig.~\ref{fig:ROI}), due to previous treatments. In such cases a ROI comprehensive of both the uptaking and the interstitial regions was chosen. 

For each of the $N_s$ slices, the effective number of voxels ($N^v_i$),  the mean value $\mu_i$ ($i=1,...,N_{s}$), and the standard deviation $\sigma_i$ of the voxel specific activity was measured. From this information the mean value of the tumor specific activity $\mu$ and the corresponding uncertainty $\sigma_\mu$  was estimated with a weighted average
\begin{equation}
\label{eq:mu}
\mu=\frac{\sum_i w_i \mu_i} {\sum_i w_i}  \hspace{3cm} \sigma_\mu =\frac{1}{\sqrt{\sum_i w_i}}
\end{equation}
where
$w_i= \frac{N^v_i}{\sigma_i^2} $.

\begin{figure}[bth!]
\begin{center}
\includegraphics [width = 174 mm] {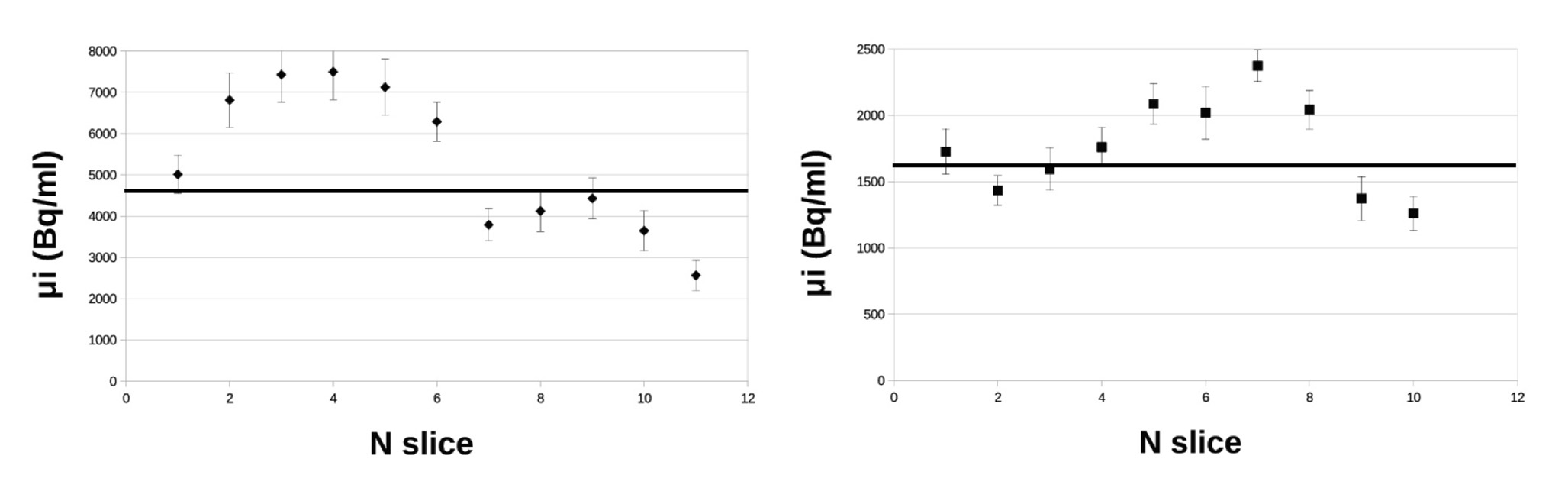}
\caption{Example of  the $\mu_i$ estimated in the different slices in the case of a meningioma (left) and glioma (right).}
\label{fig:slices}
\end{center}
\end{figure}

Fig.~\ref{fig:slices} shows typical values of $\mu_i$ for different slices and  their average. Each slice in the DICOM images is 3.27~mm thick. The effects on the margins of the tumor are masked by the spatial resolution of $\approx$3~mm. Such resolution causes a systematic underestimate of the activity concentration, effect known as partial volume effect  (PVE). We have conservatively avoided to correct for it.

In order to take into account the difference in administered activity of tracer injected to the patients, the standardized uptake value (SUV) was estimated by normalizing the measured $\mu$ by the administered activity per unit of mass of the patient, after having rescaled this activity at the time of the  PET scan:
\begin{equation}
\label{eq:SUV}
{\rm SUV}=\frac{\mu~W}{A_{adm}~e^{-0.693\Delta t_{PET}/T^{1/2}_{Ga}}}
\end{equation}
where $A_{adm}$ is the administered activity, $W$  mass of the patient, $\Delta t_{PET}$ is the time elapsed between the administration and the PET scan, and $T^{1/2}_{Ga}$ is the $^{68}$Ga half-life.

As far as the healthy tissue is concerned, several ROIs on several slices of the scan were chosen. Since the $\beta^-$ radiation is local, the ROIs were chosen, by means of the information from the CT, close to the tumor margins, as shown in Fig.~\ref{fig:ROI} (right). %Also 
In this case, a weighted average was used to evaluate $\mu_{NT}$, $\sigma_{NT}$ and the corresponding $SUV_{NT}$. 

The TNR was estimated as the ratio between $\mu$ and $\mu_{NT}$.

\begin{figure}[bth!]
\begin{center}
\includegraphics [width = 174 mm] {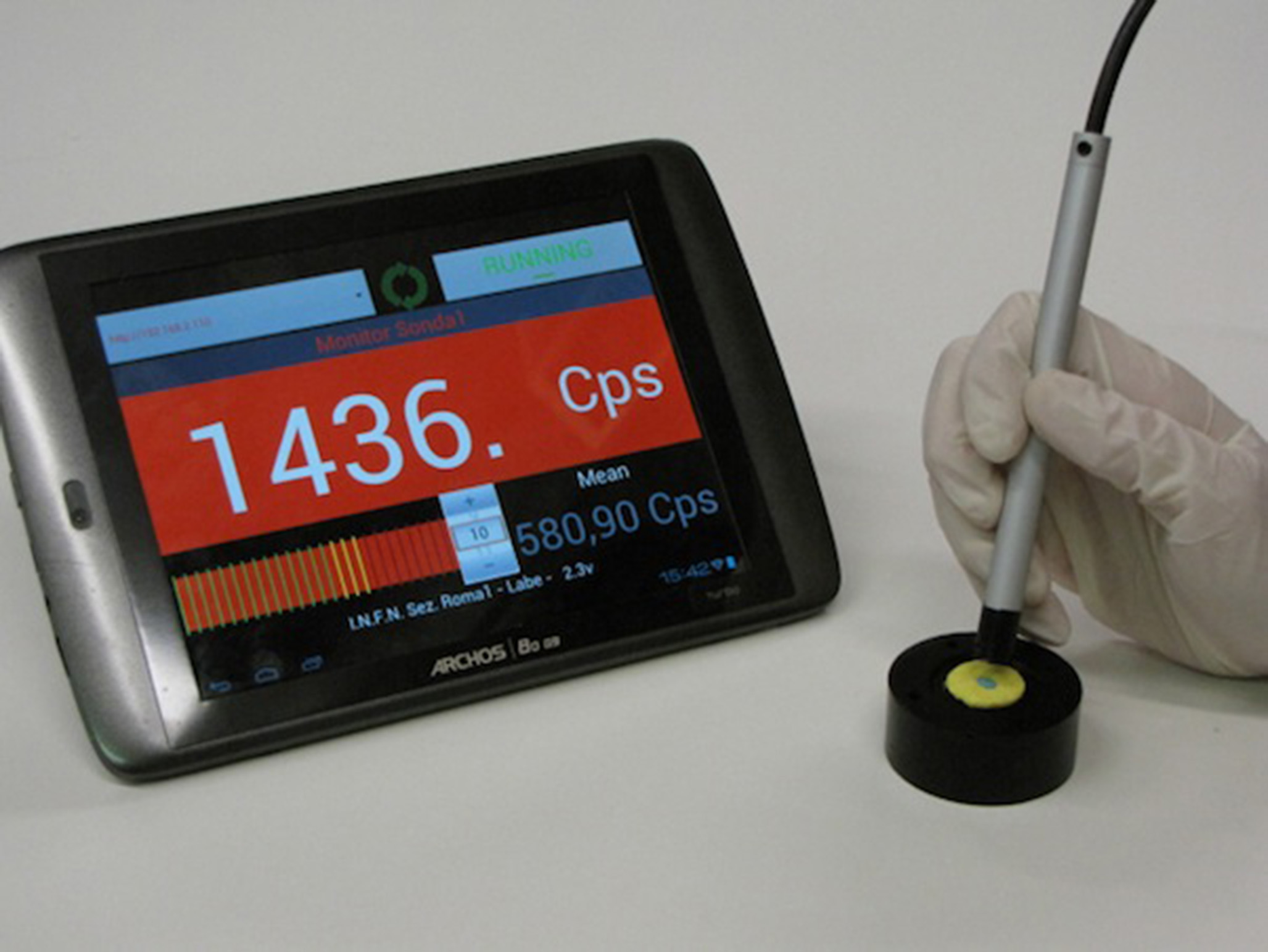}
\caption{Image of the probe prototype. }
\label{fig:probe}
\end{center}
\end{figure}

{\color{red} 
\subsection*{The $\beta^-$ Probe}

A critical element in the development of the proposed RGS method is the $\beta^-$ probe, where the low penetration power of the $\beta^-$ radiation can be exploited by reducing the size of the lateral shielding. Furthermore, we have found that we could minimize the sensitivity to $\gamma$ radiation by using an organic scintillator with low density but high light yield (p-terphenyl\mycite{pterf}). The prototype used as reference in this study has a sensitive cylinder of p-terphenyl with a radius of 2.55~mm and depth of 3~mm (Fig.~\ref{fig:probe}). 
To maximize the accuracy on the direction of the incoming radiation, the sensitive region is screened by  3~mm of PVC. The scintillation light  is transported to a photo-multiplier tube (PMT, Hamamatsu H10721-210 in the prototype) through four optical fibers. 

}% end of RED
\subsection*{Expected Performances of RGS With $\beta^-$ Decays}

From the measured specific activities we also extrapolated a prediction on the potentialities of the RGS with $\beta^-$ decays. As previously mentioned and detailed in Ref.\mycite{SciRep}, this technique exploits the fact that the tumor has a larger uptake of DOTATOC than the healthy tissues: the patient scheduled for surgery   is injected, several hours in advance, with $^{90}$Y-DOTATOC. From the DICOM images, properly accounting for the radio-nuclide half-lives and assuming that the radio-tracer is administered $\Delta t_{surg}=12$ hours before the intervention, we can estimate the expected specific activity  in tumor ($\mu^{ref}$) and non-tumor ($\mu^{ref}_{NT}$) at the time of the surgery for each patient:
\begin{equation}
\mu^{ref}_{(NT)}=SUV_{(NT)}A_{ref} e^{-0.693\Delta t_{sur}/T^{1/2}_{Y}}
\end{equation}
 where $T^{1/2}_{Y}$ is the $^{90}$Y half-life. $A_{ref}=3$~MBq/kg is a  reference administered activity per unit mass, chosen as the one injected for a typical PET scan.

During surgery and after the bulk removal, the surgeon explores the surgical site with a $\beta^-$ probe to check the completeness of the resection.The probe will respond to the presence of the $\beta^-$ radiation with a signal whose rate ($\nu$) will depend on the presence of radionuclides, their spatial distribution and the probe characteristics. When probing for possible tumor residuals, a threshold on the number of counts in a given time interval will have to be set to discriminate between healthy and lesion tissues. 

To judge the feasibility of this approach, the measured specific activities $\mu^{ref}_{(NT)}$ need to be converted in the corresponding probe rates $\nu_{(NT)}$. To this aim, we performed a full simulation, including all interactions of particles with matter, with the FLUKA\mycite{FLUKA,FLUKA2} Monte Carlo software. In this simulation the background was represented by an extended  region with a specific activity $\mu^{ref}_{NT}$ of $^{90}$Y, while the tumor residual used as benchmark is a cylinder with a radius of 3~mm and an height of 3.5~mm for a total volume of 0.1~ml. These are indeed the dimension of a typical residual that has to be identified by the probe. The tumor region is assumed to have a specific activity $\mu^{ref}$ of $^{90}$Y. 

 The FLUKA software,  fully simulating the nuclear decays and the interactions of the radiation with the tissues, was used to estimate the energy deposited inside the active volume of the probe, described above, by the emitted radiation. The conversion factor between the deposited energy and the electronic signal issued by the probe  was determined in laboratory tests with  a point-like source of $^{90}$Sr of known activity.
 When the estimated electronic signal overcomes a threshold, that was determined in the same laboratory tests, the probe is considered to have issued a count. 
 
 From this simulation we are then capable to extract the signal rates that we expect to measure with a realistic $\beta^-$ probe in presence of a tumor residual ($\nu$) or in presence of only healthy tissue ($\nu_{NT}$). Such estimates, like $\mu$ and $\mu_{NT}$ that are used as input, depend on the administered activity of tracer.

From the measured detector rates we can estimate the time needed by the probe to identify a tumor residual of 0.1~ml with a given probability of false positives (FP) or false negatives (FN). For a fixed acquisition time of the probe ($t_{probe}$) the FP and FN are  defined  as:
\begin{eqnarray}
FP&=&1-\sum_{N=0}^{N_{thr}-1}{\cal{P}}_{\nu_{NT}t_{probe}}(N)\\
FN&=&\sum_{N=0}^{N_{thr}-1}{\cal{P}}_{\nu t_{probe}}(N) .
\end{eqnarray}
where ${\cal{P}}_{\mu}(N)$ indicates the Poisson probability to have $N$ if the mean is $\mu$ and $N_{thr}$ is the threshold in counts that we expect to set on the probe signal.

The minimum time that a surgeon needs to spend on a sample to evaluate whether it is healthy or not, $t_{probe}^{min}$, is determined by finding the minimal value of $t_{probe}$ for which there exists a value of $N_{thr}$ such that FN$<5\%$ and FP$< 1\%$. 

RGS can be practical only if, when administering the reference activity  (3~MBq/kg), the time $t_{probe}^{min}$ is not significantly longer than 1~s, a reasonable time lapse in the surgical environment. Otherwise, an increase of activity would be needed. On the contrary, if  $t_{probe}^{min}$ is shorter than 1~s  there are margins to reduce the administered activity. We therefore also calculated which is the minimum activity that needs to be administered ($A_{min}$) 12 hours before surgery in order to achieve FN$<5\%$ and FP$< 1\%$ in $t_{probe}^{min}=1$~s.

It is to be noted that the scaling between activities is performed neglecting the biological wash-out of the organs. Nevertheless, wash-out is faster on the healthy tissues than in the tumor, where the tracer is bound. Tumors show typically a constant or increasing phase in the first day after administration and a wash-out after 24 h\mycite{timeUp,timeUp2}.  Since the surgery, compared to  PET scan, takes place after a longer time has elapsed, the TNR was most likely underestimated and the conclusions of this study are conservative.

\section*{RESULTS}
{\color{red}
The results of the measurements on  patients affected by meningioma and HGG are in Tabs.~\ref{tab:meningioma} and~\ref{tab:glioblastoma}. They report the 
weight of each patient (W), the administered activity in the PET with Ga ($A_{adm}$), the signal and non-tumor rate expected on the probe ($\nu$ and  $\nu_{NT}$ respectively) in each lesion, the time $t_{probe}^{min}$ needed to identify a 0.1 ml residual and the minimum activity that needs to be administered to have  $t_{probe}^{min}=1$~s ($A^{min}_{1s}$), and the diagnosis for the HGG patients considered. The last column indicates whether the patient has already undergone surgery (S), radiotherapy (RT), chemotherapy (CT), Bevacizumab (B), Immunotherapy (I) or PRRT.
  $\nu$,  $\nu_{NT}$, $t_{probe}^{min}$, and  $A^{min}_{1s}$ are computed assuming that between the administration of $^{90}$Y and the surgery 12 hours elapse, while, when estimating the first three quantities, it is assumed that the reference activity $A_{ref}=3$~MBq/kg is administered. 
  }% end of red
\subsection*{Meningioma Patients}
\begin{figure}[bth!]
\begin{center}
\includegraphics [width = 174 mm]{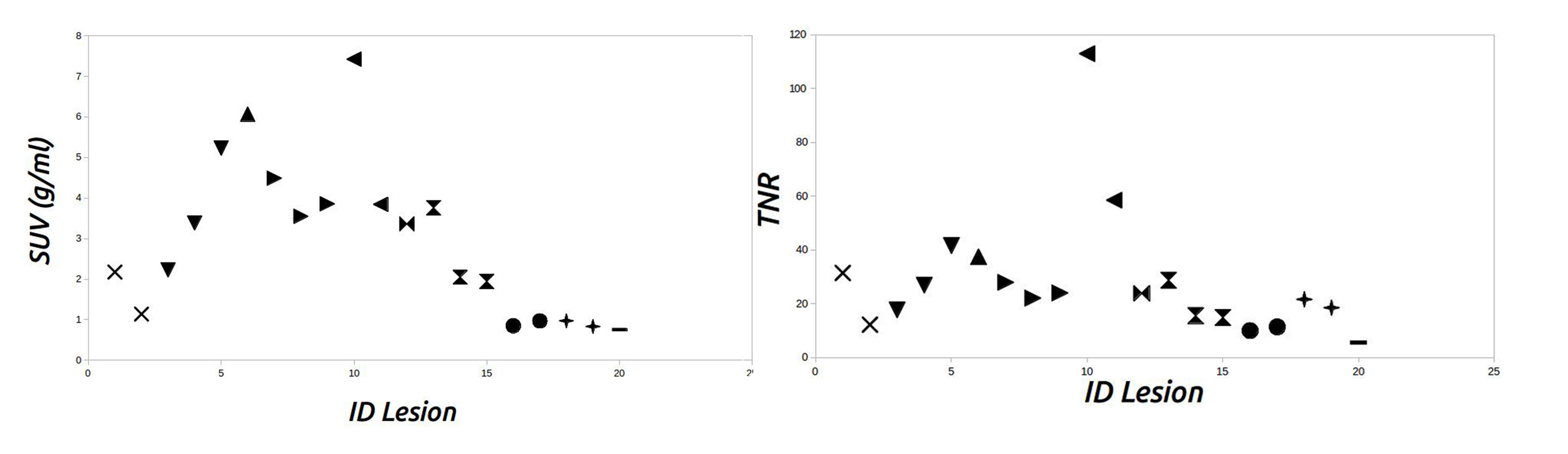}
\caption{Measured SUV (left) and TNR (right) on all studied meningioma lesions. Lesions from the same patient share the same symbols.}
\label{fig:mu_men}
\end{center}
\end{figure}

The observed SUV and TNR on the meningioma patients is shown in Fig.~\ref{fig:mu_men}.  About 70\% of the lesions have a SUV greater than 2~g/ml, i.e. administering 3~MBq/kg would  yield a specific activity greater than 6~kBq/ml. The rest of the patients have  a SUV which is half this value. In almost all cases the TNR is greater than 10, most of the times above 20.

In terms of feasibility of the RGS with $\beta^-$ emitters, Tab.~\ref{tab:meningioma} reports for each patient the signal and NT rate expected on the probe, the time $t_{probe}^{min}$ needed to identify a 0.1~ml residual if an activity $A_{ref}=3$~MBq/kg is administered and the minimum activity, scaled to the time of the surgery, that needs to be administered to have  $t_{probe}^{min}=1$~s. 

\subsection*{HGG Patients}
\begin{figure}[bth!]
\begin{center}
\includegraphics [width = 174 mm] {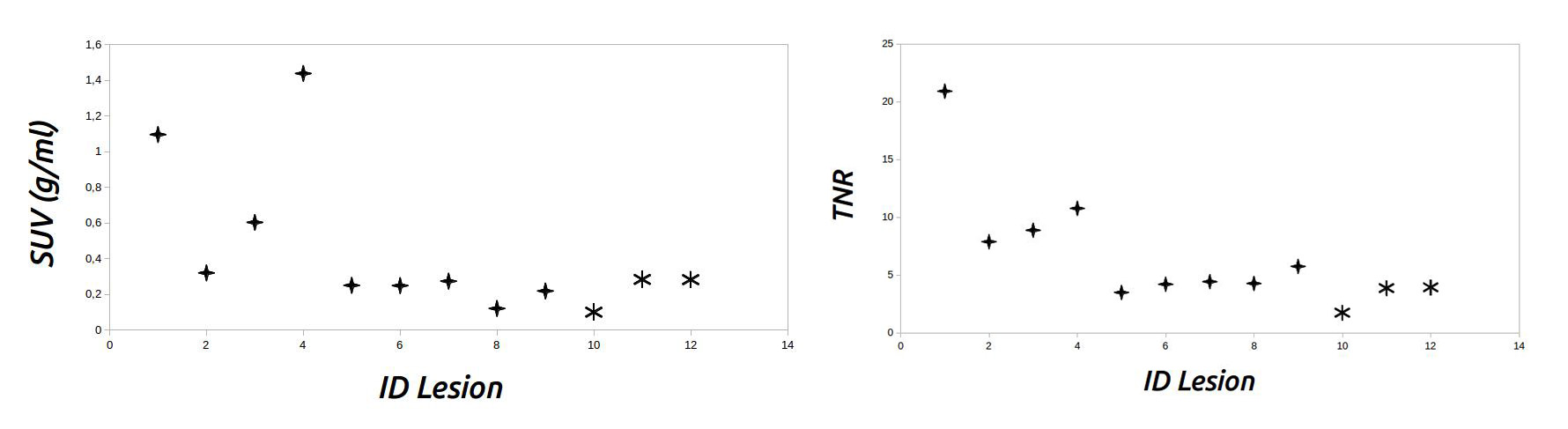}
\caption{Measured SUV (left) and TNR (right) on all studied HGG lesions.}
\label{fig:mu_gli}
\end{center}
\end{figure}

The  measured SUV and TNR on the HGG patients is shown in Fig.~\ref{fig:mu_gli}.  About 60\% of the lesions have a SUV around 0.2~g/ml, i.e. administering 3~MBq/kg would  yield a specific activity about 0.6~kBq/ml.  The TNR is always at least four, and twice as high in one third of the cases. The only case where the uptake is marginal is the only oligodendrioma in the sample.

With such indications we have calculated the information relevant for the RGS with $\beta^-$ decays (see Tab.~\ref{tab:glioblastoma}) and found that by administering 3MBq/kg the probe requires in a significant fraction of the cases about 5-6~s to discriminate between lesion and healthy tissue. In such cases, to reduce the probing time to 1~s, a therapeutical activity of 20MBq/kg would have to be administered. 

\section*{DISCUSSION}
{\color{red}
The novel RGS technique under development relies on the possibility to deliver efficiently and selectively a $\beta^-$ emitter. To this aim, this paper studies the uptake of $^{90}$Y-DOTATOC in two cases of interest, meningioma and high grade glioma.

Meningioma is known to have high receptivity to somatostatine analogues and therefore it is an ideal test bench for the proposed technique. Nonetheless, it is important to assess the impact of such large receptivity on the protocol of the RGS. In particular there needs to be assessed whether there are meningioma patients that do not express enough receptors for the RGS to be effective and which is the lowest activity that needs to be administered for the technique to be effective, in order to minimize the exposure of the patient and of the medical personnel.
The results show that in the case of meningioma only one patient out of 11 did not express enough receptors to allow for a detection of a residual of 0.1~ml in less than 1s, nonetheless also in this case the probing time is below 2~s. We can therefore conclude that the proposed RGS can be applied to all patients. Furthermore,  in 60\% of the patients the technique could be effective administering only 0.5 MBq/kg.
It is finally to be noted that the Monte Carlo approach developed here will be part of the protocol to determine the activity of $^{90}$Y-DOTATOC to administer to patients, since a PET scan with $^{68}$Ga-DOTATOC will be acquired prior to surgery.

As far as HGG patients are concerned, the literature reported a limited receptivity to DOTATOC. Nonetheless, it was important to ascertain whether the uptake and the TNR were sufficient for a diagnostic use.
the uptake was significantly less, but nonetheless the TNR was in most of the cases larger than four, even with a conservative estimate. This implies that, in case of HGG, the RGS with $\beta^-$ decays is expected to be reliable if each possible residual is examined for at most 5-6 seconds, since the activities needed to reach the required sensitivity in 1s  are too large. Alternatively, to reduce the measurement time, there are still margins to improve the sensitivity of the probe: more sensitive devices can be exploited in the scintillation light collection and crystals with a larger light yield can be considered, at the expense of a more challenging engineering of the probe. 
It is also to be noted that the estimates are conservative because  the region between areas with a good uptake is also included in the ROI since it cannot be established whether it is  still tumor or not (see Fig.~\ref{fig:ROI}).
}% end of red
\section*{CONCLUSION}

We estimated the uptake of DOTATOC in meningioma and high grade glioma in the context of the feasibility study of a novel RGS technique exploiting $\beta^-$ decays. The results showed that in the case of meningioma the uptake is so marked that the technique can work even administering activities much smaller than the one needed for PET scans. Such minimum required activity can be computed on a patient-by-patient basis. The proposed technique can also work for high grade glioma, although the limited receptivity requires a longer probing time with the current version of the probe.

\newpage
\section*{Disclosure}
F.B., F.C., E.D.L., M.M., I.M., V.P., L.P., A.Sa., A.Sc., C.V. and R.F are listed as inventors on an Italian patent application (RM2013A000053) entitled "Utilizzo di radiazione beta- per la identificazione intraoperatoria di residui tumorali e la corrispondente sonda di rivelazione" dealing with the implementation of an intra-operative beta- probe for radio-guided surgery according to the results presented in this paper. The same authors are also inventors in the PCT patent application (PCT/IT2014/000025) entitled "Intraoperative detection of tumor residues using beta- radiation and corresponding probes" covering the method and the instruments described in this paper.

\newpage

%\newpage
%\newpage
 
%\newpage

%\newpage

\begin{table}[!bth]
 \begin{center}
  \caption{Results on meningioma patients, one line per lesion. The meaning of the symbols is in the text.
  \label{tab:meningioma}}
  \begin{tabular}{| l || c | c | c || c | c | c | l|l || l|}
    \hline
    \hline
     Patient&  & W & $A_{adm}$ & $\nu$ & $\nu_{NT}$  & $t_{probe}^{min}$  & $A^{min}_{1s}$& Diagnosis &Previous   \\ 
ID & $N_{les.}$ &  (kg) & (MBq)  & (Hz) &  (Hz) &  (s) & (MBq/kg) &   &Treatment \\  \hline
     M01& 1& 63& 220& 32.2& 1.9& 0.2& 0.7 & atypical & S\\ \hline
     M02& 1& 80& 160& 17.6& 2.6& 0.6& 1.9 & atypical&S/RT/PRRT\\ \hline
     M03& 3& 95& 305& 33.7&	3.5& 0.3& 0.9 & likely atypical & S/RT\\ \hline
     & & &                    & 50.3& 3.5&	0.3& 0.5 &  & \\ \hline
     & & &                    & 76.8& 3.5&	0.1& 0.3 &  &\\ \hline
     M04& 1& 48& 200& 89.4&	4.5& 0.1& 0.2 & atypical & S/RT/CT\\ \hline
     M05& 3& 57& 130& 66.7&	4.4& 0.2& 0.3 & relapse & S/RT/CT/PRRT\\ \hline
     & & &                    & 53.2& 4.4& 0.2& 0.5 &  &\\ \hline
     & & &                    & 57.6& 4.4& 0.2& 0.4 &  &\\ \hline
     M06& 2& 90& 145& 107.6& 1.8& 0.1& 0.1 & unknown & PRRT \\ \hline
     & & &                    & 56.1& 1.8& 0.2& 0.4 &  &\\ \hline
     M07& 1& 74& 237& 50.2& 3.9& 0.2& 0.5 & anaplastic& S/RT\\ \hline
     M08& 3& 105& 223& 55.7& 3.6& 0.2& 0.5 & atypical&S/RT\\ \hline
     & & &                    & 31.2& 3.6& 0.2& 0.9 &  &\\ \hline
     & & &                    & 29.6& 3.6& 0.4& 0.9 &  &\\ \hline
     M09& 2& 48& 145& 13.4& 2.4& 0.9& 2.7 & atypical &S/RT\\ \hline
     & & &                    & 15.1& 2.4& 0.7& 2.5 &  &\\ \hline
     M10&1& 70& 240& 14.6& 1.2&	0.6& 1.8 & atypical &S/RT \\ \hline
     & & &                    & 12.6& 1.2& 0.8& 1.9 &  &\\ \hline
     M11&1& 75& 220& 12.7& 3.8&	1.6& 5.0 & atypical & unknown \\ \hline
     \hline
       \end{tabular}
 \end{center}
\end{table}

\newpage

\begin{table}[!bth]
 \begin{center}
  \caption{Results on HGG patients. The meaning of the symbols is in the text.
  \label{tab:glioblastoma}}
  \begin{tabular}{| l | c | c  || c | c | c | l | l || l|}
    \hline
    \hline
     Patient  & W & $A_{adm}$ & $\nu$ & $\nu_{NT}$  & $t_{probe}^{min}$  & $A^{min}_{1s}$& Diagnosis  &Previous  \\ 
ID &  (kg) & (MBq)  & (Hz) &  (Hz) &  (s) & (MBq/kg) & & Treatment   \\  \hline
     G01 & 97 & 246 & 16.5 & 1.4 & 0.5 & 1.5 & HGG&S/RT/CT/PRRT\\ \hline
     G02 & 68 & 223 & 5.2 & 1.1 & 2.6 & 8.5 & HGG& RT/CT/B\\ \hline
     G03 & 80 & 152 & 9.6 & 1.9 & 1.4 & 4.3 & HGG&S/RT/CT\\ \hline
     G04 & 93 & 198 & 22.4 & 3.7 & 0.6 & 1.8 & HGG&S/RT/CT/PRRT\\ \hline
     G05 & 90 & 192 & 4.6 & 2.0 & 7.4 & 23.6 & HGG&S/RT/CT/PRRT\\ \hline
     G06 & 60 & 185 & 4.4 & 1.6 & 5.8 & 20.0 & HGG&S/RT/CT\\ \hline
     G07 & 63 & 194 & 4.8 & 1.7 & 5.1 & 17.6 & HGG&S/RT/CT\\ \hline
     G08 & 70 & 266 & 2.1 & 0.8 & - & 40.0 & HGG&RT/CT\\ \hline
     G09 & 85 & 255 & 3.7 & 1.1 & 5.3 & 17.6 & HGG & S/RT/CT\\ \hline
     G10 & 80 & 224 & 2.2 & 1.6 & - & - & oligodendroglioma & S/RT/CT/I\\ \hline
     G11 & 70 & 234 & 5.1 & 2.0 & 5.5 & 18.8 & HGG & RT/CT\\ \hline
     G12 & 15 & 38 & 5.0 & 2.0 & 5.9 & 18.8 & pontine glioma & RT/CT/PRRT\\ \hline
  \end{tabular}
 \end{center}
\end{table}
\end{document}